\documentclass[global,twocolumn]{svjour}
\usepackage{graphics}
\usepackage{graphicx}
\usepackage{amsmath}
\usepackage{amsfonts}
\usepackage{amssymb}
\usepackage{pstricks}
\usepackage{psfrag}
\usepackage{changes}
\usepackage[ansinew]{inputenc}
\usepackage{epstopdf, epsfig}
\begin{document}
\title{Hydrodynamic interaction of a self-propelling particle with a wall}
\subtitle{Comparison between an active Janus particle and a squirmer model}
\author{Zaiyi Shen\inst{1}, Alois W\"urger\inst{1} \and Juho S. Lintuvuori\inst{1}
}                     
\offprints{juho.lintuvuori@u-bordeaux.fr}          
\institute{Univ. Bordeaux, CNRS, LOMA, UMR 5798, F-33405 Talence, France } 
%
\date{Received: date / Revised version: date}
%
\abstract{
  Using lattice Boltzmann simulations we study the hydrodynamics of an active spherical particle near a no-slip wall. We develop a computational model for an active Janus particle, by considering different and independent mobilities on the two hemispheres and compare the behaviour to a standard squirmer model.
  We show that the topology of the far-field hydrodynamic nature of the active Janus particle is similar to the standard squirmer model, but in the near-field the hydrodynamics differ. In order to study how the near-field effects affect the interaction between the particle and a flat wall, we compare the behaviour of a Janus swimmer and a squirmer near a no-slip surface via extensive numerical simulations. Our results show generally a good agreement between these two models, but they reveal some key differences especially with low magnitudes of the squirming parameter $\beta$. Notably the affinity of the particles to be trapped at a surface is increased for the active Janus particles when compared to standard squirmers. Finally we find that when the particle is trapped on the surface, the velocity parallel to the surface exceeds the bulk swimming speed and scales linearly with $|\beta|$.
%
} 
\maketitle
\section{Introduction}
\label{intro}
Artificial microswimmers have become an important tool to study the structures and dynamics of motile micro-organisms in a laboratory~\cite{Zottl16,ReviewBechinger}. A typical example of artificial swimmer is provided by an active Janus particle: a spherical micrometer sized colloidal particles rendered motile by local gradients ({\it} e.g. chemical or thermal)~\cite{ReviewBechinger}. These local gradients give a rise to a phoretic slip velocity tangential to particle surface, resulting a squirming motion of the particle~\cite{ReviewBechinger}. A popular theoretical model for describing the squirming motion was introduced by Lighthill~\cite{Lighthill52} (so called squirmer model). In this model, a continuous slip velocity is assigned to the particle surface, leading to a time-independent squirming motion~\cite{Magar03}. The squirmer model has been instrumental in theoretical and simulation studies of the hydrodynamics of spherical self-propelling particles~\cite{ignacio1,ignacio2,ishimoto13,zottl14,li14,lintuvuori16,lintuvuori17}. The hydrodynamic nature of the motion is given by the squirming parameter $\beta$, where $\beta<0$ corresponds to pushers and $\beta > 0$ to pullers, respectively.

Both experimental and simulation studies have shown that the swimmers have an affinity to accumulate near surfaces~\cite{ishimoto13,zottl14,li14,brown15,das15,lintuvuori16}. Experiments have shown that the synthetic swimmers can be trapped by a colloidal crystal~\cite{brown15} and solid geometries can be used to guide the particles~\cite{das15,Sim16}. Specifically simulations~\cite{ignacio2,li14,lintuvuori16} and detailed theoretical calculations~\cite{ishimoto13} have shown that the trapping of the squirmers by a no-slip wall is strongly dependent of the value of $\beta$: typically $|\beta|\gtrsim 4$ is required~\cite{ishimoto13}.
  Also specific surface interactions can influence the observed dynamics. For example, including a short range repulsion near the wall, can lead to both decaying or periodic cyclic motion near the surface~\cite{ignacio2,li14,lintuvuori16}.

In the experimental realisation of artificial microswimmers based on two faced Janus particles, the two hemispheres are expected to have different interaction with the medium. An example of this is provided by differing catalytic rates ({\it e.g.} reaction with hydrogen peroxide H$_2$O$_2$ in the case of chemical swimmers~\cite{brown14,brown15,ebbens12b,sabass10,ebbens12,ebbens14,wang15,das15}) or for example different heat conduction properties leading to a temperature gradient across the particle surface for thermophoretic colloids~\cite{volpe11,Buttinoni12}. Thus it is natural to assume that the slip velocity would have a discontinuity across the equator of the particle. This is in stark contrast with the squirmer model, where the tangential slip velocity varies continuously over the the particle surface~\cite{Magar03}. While the far-field topology of the flow field is expected to only depend on the nature of the swimmer ({\it e.g.} pusher versus puller), the near-field effects could change dramatically, when the two different mobilities on the opposite hemispheres are taken into account. The near-field effects have been shown to be important in determining the behaviour of the swimmer near a no-slip surface~\cite{lintuvuori16}. It should be noted that recent theoretical and simulation work suggest that phoretic interactions can dominate particle-particle interactions~\cite{benno17,IgnacioArxiv}, and can lead to the experimentally observed clustering of synthetic swimmers~\cite{theurkauff12,palacci13,buttinoni13}. These interactions could also have an effect on the surface accumulation of synthetic swimmers.

In this work we develop a hydrodynamic model for a Janus swimmer in the framework of lattice Boltzmann (LB) simulations. We do not explicitly deal with phoretic interactions, but instead build a model for a Janus swimmer by considering a discontinuous slip velocity over the particle surface to take into account the varying mobilities in the opposing hemispheres. The model is based on LB implementation of squirmers~\cite{ignacio1,ignacio2}, but instead of continuous varying slip velocity, we assign two different mobilities on the two hemispheres $u_1$ and $u_2$. To compare to the traditional squirmer model, we map the two mobilities to the squirming parameter $\beta$. Via extensive simulations we compare the hydrodynamic interactions between the active colloid and no-slip wall, for the two models: our active Janus model and classical squirmer model. Our results demonstrate that qualitatively the Janus swimmers behave in an agreement with the squirmers. However we show that for certain values of $\beta$ the Janus swimmers can exhibit multiple different modes of the cyclic motion and stronger affinity to accumulate at the surface, this is especially true to low values of $|\beta|$.

\section{Computational model}
\label{sec:1}
%
%

\begin{figure}
\includegraphics[width=0.5\textwidth]{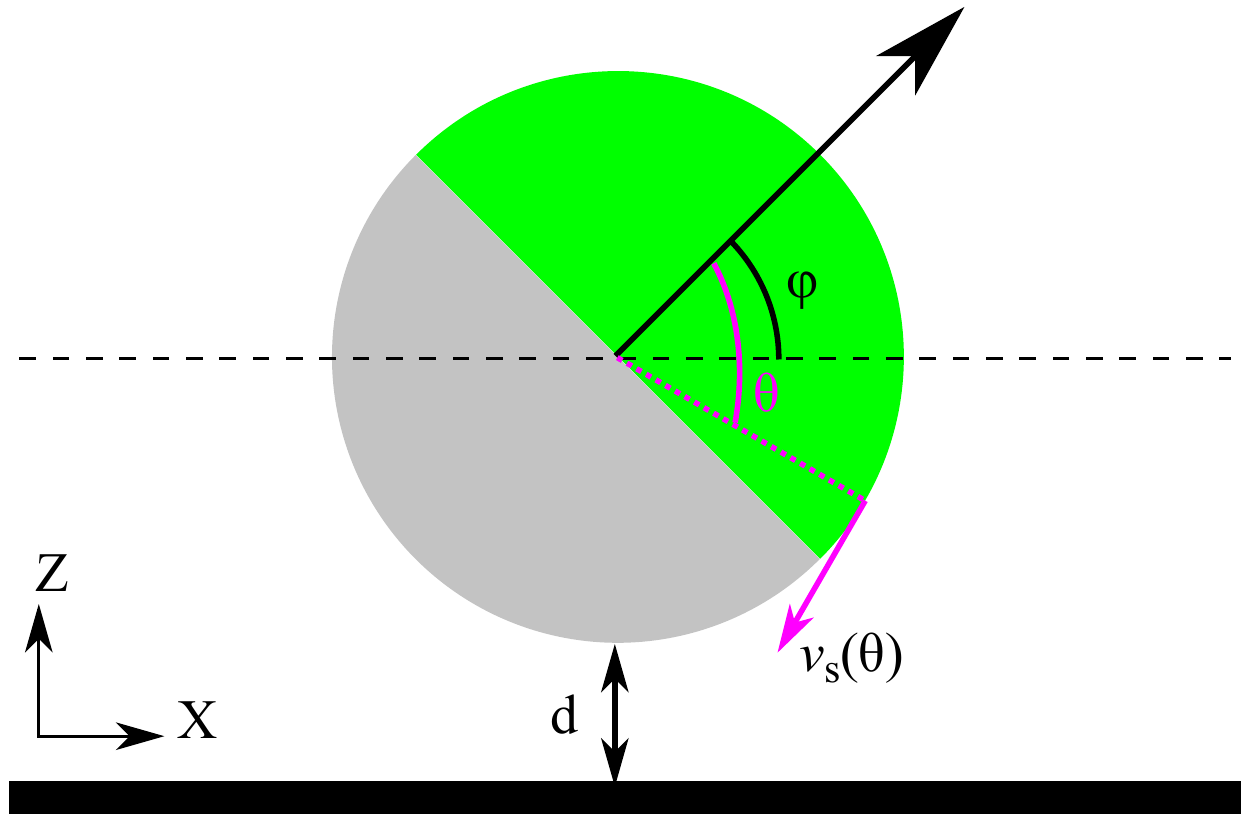}
\caption{Schematic of the microswimmer near a wall, defining the shortest distance $d$ between the particle surface and the wall as well as the angle $\varphi$ between the particle direction and the wall, used in the text. $v_s(\theta)$ is the slip velocity at the particle surface at an angle $\theta$ from the swimmer direction.}
\label{schematic}       
\end{figure}

\begin{figure*}
\includegraphics[width=1\textwidth]{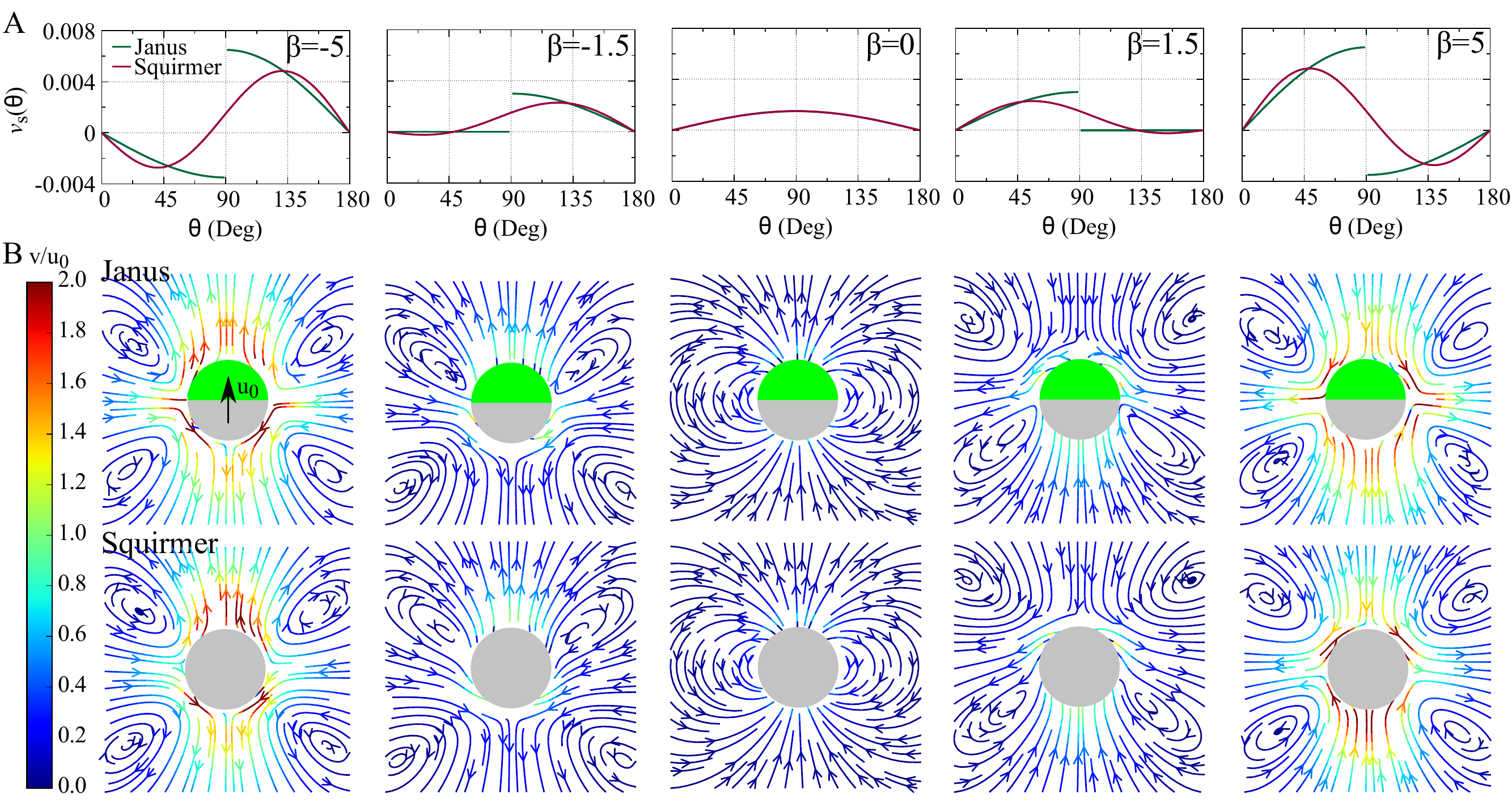}
\caption{(A) The surface velocity of an active Janus particle (green line) and a squirmer swimmer (red line) for five values of $\beta$ corresponding to a pusher ($\beta=-5, \beta=-1.5$), neutral ($\beta=0$) and puller ($\beta=1.5, \beta=5$). (B) The corresponding simulated flow fields in the bulk (lab frame): Active Janus particle (top row) and a standard squirmer (bottom row). The particle swimming direction is vertically up the page.}
\label{flowfield}       
\end{figure*}


Chemically or thermally driven Janus particles have become a widely studied class of artificial micro-swimmers. In most cases they consist of colloidal spheres with an active cap which typically covers half of their surface. The particles' activity results in a finite slip velocity $u$ which, because of different material properties takes different value on the uncoated and coated hemispheres. When defining the polar angle $\theta$ with the respect to the particle's axis (Fig.~\ref{schematic}), one has

\begin{equation}
  v_s(\theta)=\begin{cases}
  u_1\sin \theta  &  \text{for } \cos \theta \geq 0 \\ 
  u_2\sin \theta  &  \text{for } \cos \theta < 0 
\end{cases}
\label{janus}
\end{equation}
where $u_1$ and $u_2$ describe the slip velocity on the upper and lower hemispheres, respectively.   The factor $\sin\theta$ is due to the spherical geometry; more intricate dependencies may arise from the particle's material properties \cite{bickel13} or from its activity \cite{Bic14}.

Expanding the step function in Eq. (\ref{janus}) in terms of Legendre polynomials $P_n(\cos\theta)$, one has
\begin{equation}
v_s(\theta) = \sin\theta \sum_{n=0}^\infty p_n P_n(\cos\theta).
\end{equation}
The even coefficients vanish except for the first one,  $p_0=\frac{1}{2}(u_1 + u_2)$, whereas the odd ones are finite,
$p_1=\frac{1}{2}(u_1 - u_2)$, $p_3=-\frac{1}{8}(u_1 - u_2)$,... The first two terms of this series correspond to the standard squirmer model,
\begin{equation}
  v_s(\theta) =  \frac{3}{2}  u_0 \sin\theta (1 + \beta \cos\theta ),
  \label{squirmer}
\end{equation}
where $u_0\equiv\tfrac{2}{3}p_0$ is the unperturbed bulk swimming speed and $\beta\equiv p_1/p_0$ is the squirming parameter, which can be used to distinguish between $\beta>0$ pullers and $\beta < 0$ pushers.
The parameters between the Janus particle and squirmer models are related through
\begin{equation}
u_0 =  \frac{u_1 + u_2}{3}  , \;\;\;  \beta =  \frac{3}{2}  \frac{u_1 - u_2}{u_1 + u_2},
\end {equation}
implying $u_{1,2} = u_0(\frac{3}{2} \pm \beta)$.

\if{
To model the active Janus particle we consider the simplest case which has a following surface slip velocity

where the $u_1$ and $u_2$ are the independent mobilities of the two hemispheres and $\theta$ is the polar angle (Fig. 1).
These give a bulk swimming velocity as $u_0= (u_1+u_2)/3$.

To model the squirmer, we employ the standard form of tangential slip velocity at the particle surface~\cite{Magar03}
\begin{equation}
u(\theta)=B_1\sin \theta + B_2 \sin \theta  \cos \theta .
\label{squirmer}
 \end{equation}
This gives velocity of the particle in bulk as $u_0= 2 B_1/3$ and the squirming parameter $\beta={B_2}/{B_1}$. The $\beta$ gives the nature of the swimmer: when $\beta<0$ the swimmer is a pusher and $\beta>0$ gives a puller.

To compare the results between the two models (Janus swimmer eq.~(\ref{janus}) and a squirmer eq.~(\ref{squirmer})) we can map the ratio between the mobilities of the two hemispheres of Janus particle $(u_1-u_2)/(u_1+u_2)$ to the squirming parameter $\beta$ by identifying
\begin{equation}
B_1\sin \theta (1 + \beta \cos \theta )=\begin{cases}
u_1\sin \theta &  -\frac{\pi}{2} \leq \theta \leq \frac{\pi}{2}\\

u_2\sin \theta &  \frac{\pi}{2} < \theta < \frac{3\pi}{2}
\end{cases} .
\label{identify}
\end{equation}

Integrating both sides over $\theta$ from 0 to $\pi$, we find $B_1=(u_1+u_2)/2$. Multiplying both sides with $\cos \theta$ and integrating over $\theta$ from 0 to $\pi$, we obtain $B_1 \beta = 3(u_1-u_2)/4 $. Thus we get the relationship

\begin{equation}
\beta = \frac{3}{2} \frac{u_1-u_2}{u_1+u_2} .
\label{map}
\end{equation}
}\fi

We use lattice Boltzmann method (LBM) to simulate the squirmers. In the LBM a no-slip boundary condition at solid-fluid interface can be realised by using a bounce back on links method~\cite{ladd1,ladd2}. When considering a mobile particle, the bounce back on links needs to be modified to take into account the rotational and translational motion of the particle surface. In order to simulate the squirming motion, the boundary condition at the particle surface is modified to include the surface slip flow~\cite{ignacio1,ignacio2}. To model the active Janus particle we use the slip velocity given by  eq.~(\ref{janus}) while for squirmer standard slip flow given by eq.~(\ref{squirmer}) is used.

To stop the particles to penetrate the wall, we employ a short range repulsive potential
\begin{equation}
  V(d) = \epsilon\left(\frac{\sigma}{d}\right)^\nu
  \label{soft}
\end{equation}
which is cut-and-shifted by
\begin{equation}
  V_W(d) = V(d) - V(d_c) - (d-d_c) \frac{\partial V(d)}{\partial d}\mid_{d=d_c}
  \label{repulsion}
\end{equation}
to ensure that the potential and resulting force go smoothly to zero at the interaction range $d_c= 1.2$ in simulation units (SU) corresponding to repulsion range of $\approx 0.15R$. The $\epsilon = 0.06$ and $\sigma = 1.0$ are constant in the reduced units of energy and length, respectively. The $\nu = 12$ controls the steepness of the repulsion.

\begin{figure*}
\includegraphics[width=1\textwidth]{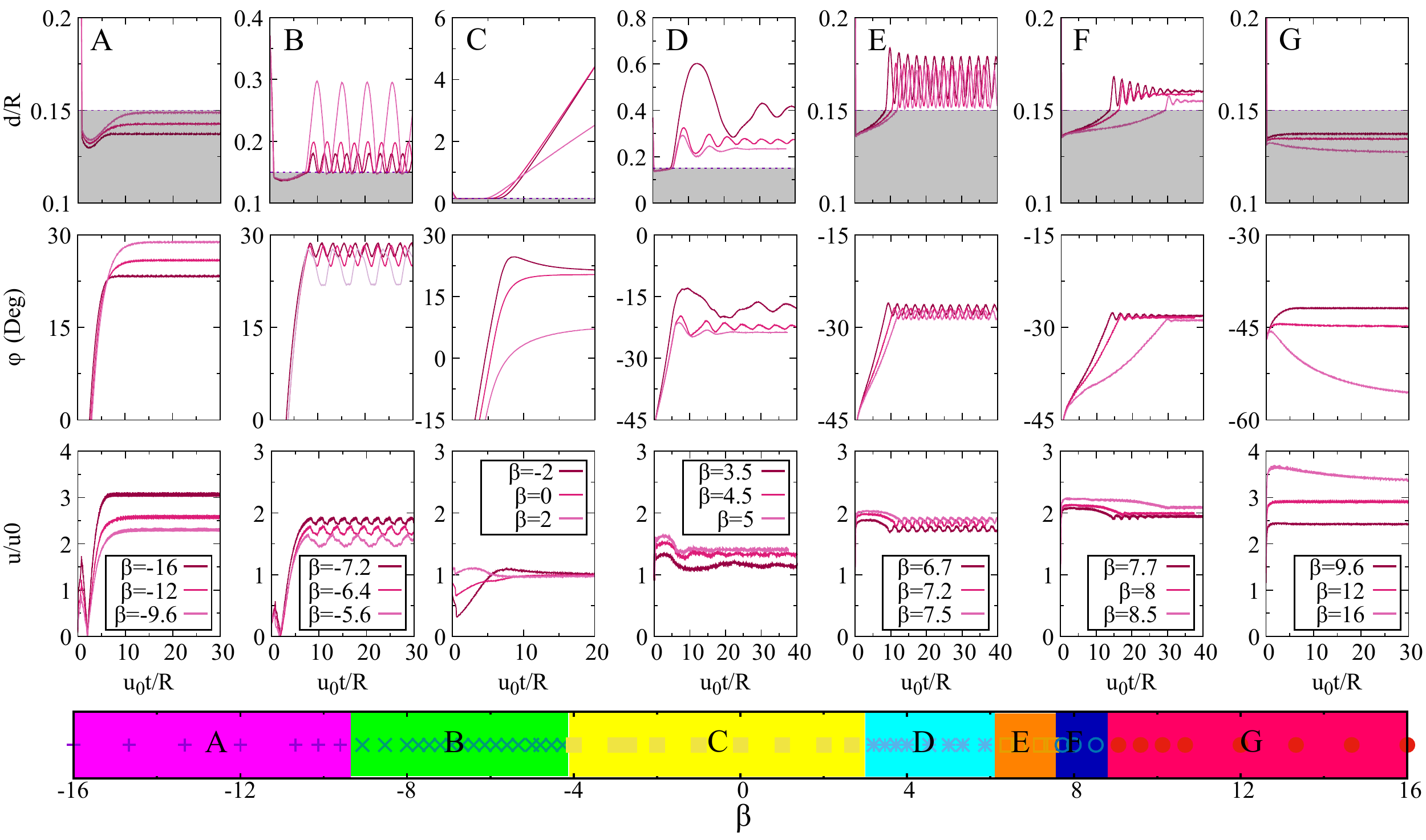}
\caption{Squirmer swimming modes (A-G): Seven different swimming modes (A-G) observed for the standard squirmer for $\beta\in[-16,+16]$ near a no-slip surface. The shortest distance between the particle surface and the wall $d(t)$ (top row), the angle $\varphi (t)$ between the swimmer orientation and the wall (2nd row) as well as the magnitude of the swimming velocity $u(t)/u_0$ (3rd row), are used to define an approximate state diagram (bottom), giving the approximate boundaries between the swimming modes: (A) strong attraction, (B) periodic swimming, (C) no trapping, (D) decaying cyclic swimming, (E) periodic swimming, (F) decaying periodic swimming and (G) strong attraction. See text for more details. (The shading in the plots of $d(t)$ (top row) corresponds to the range of the external soft repulsion.)}
\label{phase_squirmer}       
\end{figure*}

We carried out our simulations in a rectangular simulation box with the size of $96\times 96\times 96$, with a no-slip wall at $z=0.5$ and $z=96.5$ and periodic boundary conditions along $x$ and $y$ (for a schematic see Fig~\ref{schematic}) . Unless otherwise mentioned, we carried out the simulations using a particle with radius $R = 8$SU. We used a kinematic viscosity $\eta = 1/6$SU and fixed the unperturbed bulk swimming speed $u_0=10^{-3}$SU, but considered a large range of the squirming parameter $\beta\in[-16,16]$. We control our system in a Stokes regime by the small Reynolds number (Re) giving the ratio of inertial and viscous forces. Using the parameters above, we find $\text{Re}={u_0R}/{\eta}\approx 0.05$.

\section{Results}
\subsection{Flow field around a single particle in the bulk}
To model the active Janus particle, we consider a surface slip velocity with two mobilities on the opposite sides of the particle as given by the equation (\ref{janus}), while the squirmer boundary conditions are given by the equation (\ref{squirmer}).
In Fig.~\ref{flowfield} we compare both the surface slip velocity $v_s(\theta)$ and the bulk flow in lab frame, between active Janus particles and standard squirmer model, for five values of $\beta$ corresponding to the pusher ($\beta=-5,~\beta=-1.5$), neutral ($\beta = 0$) and puller ($\beta = +1.5,~\beta=+5$) swimmers. The key difference between the two types of the swimmers is the discontinuity of $v_s(\theta)$ for a Janus swimmer (green line in Fig.\ref{flowfield}(A)) while for a squirmer $v_s(\theta)$ varies continuously (red line in Fig.\ref{flowfield}(A)). This is clearly visible for the graphs for a strong $\beta = -5$ pusher and $\beta = +5$ puller (Fig.~\ref{flowfield}(A)). These correspond to the ratio of the mobilities $u_1/u_2 = -13/7$ and $u_1/u_2 = 7/13$ for $\beta = +5$ and $\beta = -5$, respectively. The $\beta = \pm 1.5$ corresponds to the special case where $u_1 = 0$ ($u_2 = 0$) for $\beta = -1.5$ pusher ($\beta = + 1.5$ puller) as shown in Fig.~\ref{flowfield}(A). For $\beta = 0$ neutral swimmer the requirement is $u_1=u_2$, thus the two models have an identical surface slip velocity $v_s(\theta)$ (middle panel in Fig.~\ref{flowfield}(A)).

Despite the large differences between the surface flow of these two models (as shown in Fig.~\ref{flowfield}(A)) the topology of the bulk flow away from the particle surface is remarkably similar between the two models. This is clearly apparent from the visualisation of the streamlines for the Janus (squirmer) swimmer as shown in top (bottom) of Fig.~\ref{flowfield}(B). These correspond to the surface slip-velocities as presented in Fig.~\ref{flowfield}(A) for a pusher ($\beta=-5,~\beta=-1.5$), neutral ($\beta = 0$) and a puller ($\beta = +1.5,~\beta=+5$). Thus in far-field these two models should have almost identical behaviour. However, for the dynamics near a no-slip wall, the near-field lubrication effects can become very important~\cite{zottl14,lintuvuori16}. Thus the differences on the surface slip velocity between these two models, could lead to a different dynamics near a boundary. In what follows the hydrodynamic interaction between these two models and a no-slip wall is analysed via lattice Boltzmann simulations.

\subsection{Different swimming modes near the surface}
\begin{figure*}
\includegraphics[width=1\textwidth]{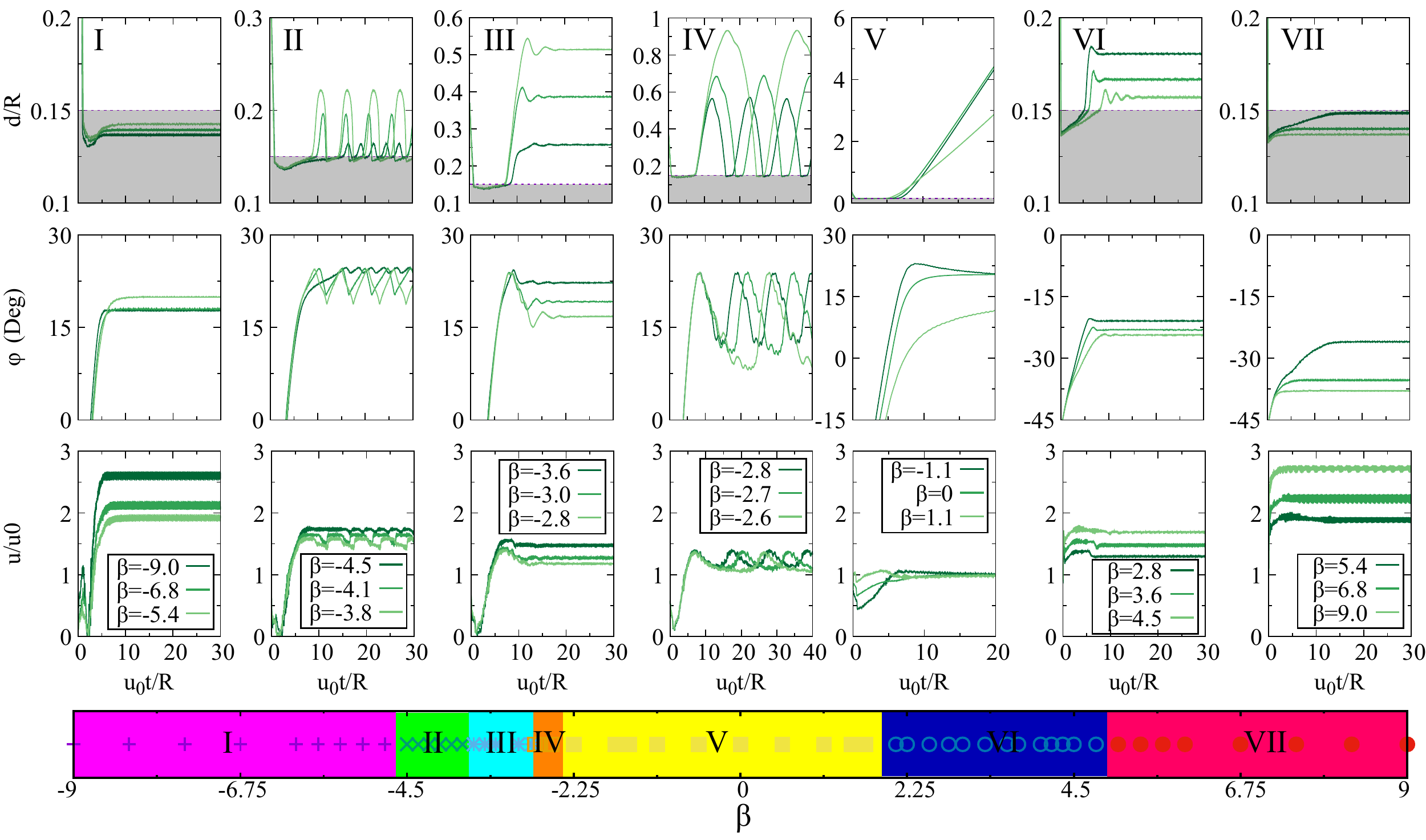}
\caption{Active Janus particle swimming modes (I-VII): Seven different swimming modes (I-VII) observed for the active Janus particles for $\beta\in[-9,+9]$ near a no-slip surface. The shortest distance between the particle surface and the wall $d(t)$ (top row), the angle $\varphi (t)$ between the swimmer orientation and the wall (2nd row) as well as the magnitude of the swimming velocity $u(t)/u_0$ (3rd row), are used to define an approximate state diagram (bottom), giving the approximate boundaries between the swimming modes: (I) strong attraction, (II) periodic swimming, (III) stable swimming near the surface, (IV) periodic swimming, (V) no trapping, (VI) stable swimming near the surface and (VI) strong attraction. See text for more details. (The shading in the plots of $d(t)$ (top row) corresponds to the range of the external soft repulsion.)}
\label{phase_janus}       
\end{figure*}

To study the dynamics of a squirmer and an active Janus particle near a no-slip surface we place the particle initially in the X-Z plane with $\varphi=-45^{\circ}$ and $d=3R/8$. To quantify the dynamic state of the swimmers, during the course of the simulation, we measure the shortest distance between the particle surface and the wall $d$, and the angle $\varphi$ between the swimmer director and the wall (see Fig.~\ref{schematic} for a schematic of the set-up). We consider both standard squirmer modeled by a surface slip velocity given by eq.~\ref{squirmer} and Janus swimmer which has two independent surface flow mobilities, leading to a discontinuous surface flow (eq.~\ref{janus}).
We map out the approximate phase diagrams for standard squirmer and a Janus swimmer near a surface. In both cases seven distinct swimming modes are observed for both the standard squirmer (A-G; Fig.~\ref{phase_squirmer}) and the Janus swimmer (I-VII; Fig.~\ref{phase_janus}).

We start by considering the standard squirmer model (eq.~\ref{squirmer}).  We explore the $\beta$ values from -16 to 16. In Fig. \ref{phase_squirmer}, the time evolution of the distance between the wall and the particle surface $d$, the angle between the particle direction and wall $\varphi$  and the absolute velocity of the particle $u$ are plotted. We observe seven distinct swimming modes, when $\beta$ is varied: After an initial re-orientation, strong pushers $\beta<-9.3$ swim near the surface while leaning to the external soft repulsion ($d<0.15R$; shaded region in the top row in Fig.~\ref{phase_squirmer}; mode A). The particle moves in a steady state with a constant $d$, $\varphi$ and $u$. The swimmer direction remains oriented away from the wall $\sim\varphi \in [23^{\circ},30^{\circ}]$, with $\varphi$ increasing with $\beta$.
For $-9.3<\beta<-4.2$, the particle undergoes a periodic motion, where it escapes slightly from the wall and returns to the repulsion range due to the reorientation by hydrodynamic torques (mode B). This is in agreement with previous observations of $\beta = -5$ pusher near a no-slip wall with an additional repulsion at the surface~\cite{lintuvuori16}. The periodic oscillation in $d$, $\varphi$ and $u$ show an increase in the period and amplitude with increasing $\beta$ (2nd column in Fig.~\ref{phase_squirmer})
However the average value of $\varphi$ and $u$ decrease with increasing $\beta$, while averaged $d$ increases.
For $-4.2<\beta<3.0$ we observe no trapping by the surface (mode C in the Fig.~\ref{phase_squirmer}). After an initial interaction and re-orientation, the particle escapes from the wall. This agrees well with the more precise theoretical calculations of a spherical squirmer near no-slip boundary~\cite{ishimoto13}, where no stable fixed points in $d,~\varphi$ space were found for this parameter range.

In the moderately strong puller regime $3.0<\beta<6.2$, the particle is again trapped by the wall, with a decaying cyclic motion (mode D in Fig.~\ref{phase_squirmer}). In a steady state the particle swims parallel to the surface with a constant $d$ and $\varphi$, while pointing towards the wall, $\varphi<0$ (D in Fig.~\ref{phase_squirmer}). This observation is in agreement with previous simulations~\cite{li14,lintuvuori16} and theoretical calculations~\cite{ishimoto13}. In this mode, the $d$ and $\varphi$ are observed to decrease when $\beta$ increases.
When the $\beta$ is increased ($6.2<\beta<7.6$; mode E in Fig.~\ref{phase_squirmer}), the particle undergoes a sustained cyclic motion. Compared to the mode D, both the amplitude and period are decreased. Further increasing the squirming parameter ($7.6<\beta<8.8$), a re-entrant behaviour of the decaying cyclic motion is observed (mode F in Fig.~\ref{phase_squirmer}).
It should be noted, that in all the cases where the cyclic motion is observed (D, E and F), in the steady state the swimmers do not interact with the external repulsion at the wall, {\it i.e.} $d(t)>0.15R$ (shaded region in the top row of Fig.~\ref{phase_squirmer}). Thus all the particle-wall interactions are purely hydrodynamic. This is opposite for very strong pullers (mode G) $\beta>8.8$. Here the particle swims along the boundary, while continuously leaning on the external repulsions $d<0.15R$ (G in Fig.~\ref{phase_squirmer}. This is similar to what was observed to strong pushers, but for pullers $d$ decreases with $\beta$, as oppose to increasing $d$ with $\beta$ as seen for very strong pushers (mode A in Fig.~\ref{phase_squirmer}.)

\begin{figure*}
\includegraphics[width=1\textwidth]{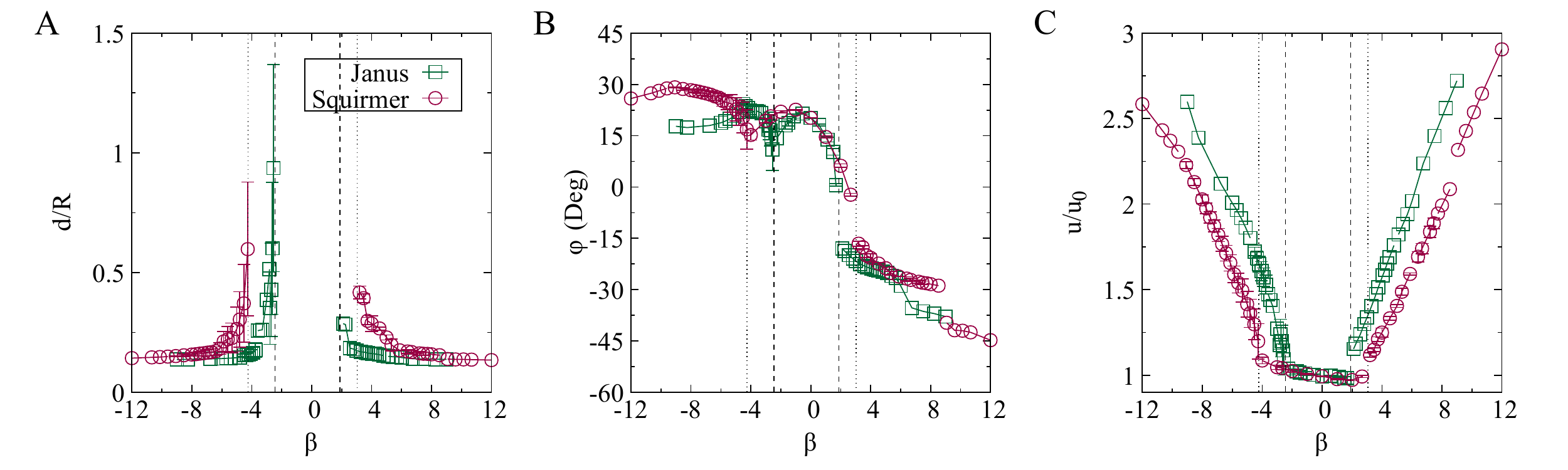}
\caption{Time averaged simulation results observed for active particles (radius $R$) for a steady state swimming near a no-slip surface. (A) shortest distance $d/R$ between the particle surface and the wall, (B) the angle $\varphi$ between the swimmer director and the wall and (C) normalised swimming speed $u/u_0$ as a function of $\beta$, for standard squirmers (open circles) and Janus swimmers (open squares). The error bars give the amplitude of the oscillations. (The vertical dotted (dashed) lines gives the boundaries of the observed trapping at surface for squirmer (Janus) swimmers.) }
\label{beta}       
\end{figure*}

  Next we turn to the active Janus particle, where the two hemispheres have different mobilities leading to discontinuous slip velocity at the particle surface (eq.~\ref{janus} and Fig. ~\ref{flowfield}(A)). In Fig.~\ref{phase_janus}, we explore the $\beta$ values from -9 to 9 for the Janus swimmer. Similarly to the standard squirmers, we observe seven different swimming modes (I-VII): (I) $\beta<-4.7$, corresponding to the mode A for a strong pusher squirmer. The particle moves along the wall, while leaning on the soft external repulsion.  Decreasing the magnitude of the $\beta$ the particle undergoes a cyclic swimming motion, escaping slightly from the wall and then returning to the repulsion regime due to the re-orientation by the hydrodynamic torques (mode II for $-4.7<\beta<-3.7$). This is similar with the mode B observed with squirmers.

  For Janus pushers with $-3.7<\beta<-2.8$ we observe a steady swimming near the boundary (mode III in Fig.~\ref{phase_janus}). Here, in a steady state the particle points away from the wall $\varphi > 0$ in agreement with the squirmer pushers. However for the Janus particle we observe that in the steady state the particle swims outside the repulsion range {\it i.e.} $d(t) > 0.15R$ (Fig.~\ref{phase_janus}; mode III), suggesting that there exists a stable fixed point in the $(d,\varphi)$ space. This is markedly different from squirmers, where the steady state dynamics for pushers only observed to undergo a cyclic motion or lean to the external repulsion.
  For  $-2.8<\beta<-2.4$, we observe a re-entrant behaviour of the periodic dynamics similar to mode II, but with larger amplitude and longer period (mode IV).

  Similarly to squirmers, we observe no trapping of the Janus swimmers by the wall for small values of $\left|\beta\right |$, as shown in Fig.~\ref{phase_janus}(V). Notably the unstable region is reduced for the Janus swimmers $-2.4<\beta<1.9$, when compared to standard squirmers $-4.2<\beta<3.0$ (modes C and V in Fig.~\ref{phase_squirmer} and Fig.~\ref{phase_janus} for Squirmer and Janus swimmers, respectively).

  For Janus pullers ($\beta > 0$) we observe two stable swimming modes: (VI) $1.9<\beta<4.9$, the particle is trapped by the no-slip surface with a well defined steady state $d$ beyond the external repulsion and pointing towards the wall, $\varphi < 0$ in agreement with squirmers.  Finally strong pullers (VII) $\beta>4.9$, are strongly trapped by the surface and in steady state they lean on the repulsion, while retaining the orientation towards the wall, similarly to strong squirmer pullers.

\subsection{Comparison between Janus swimmers and squirmers}
To allow more detailed comparison we plot in Fig.~\ref{beta}, the steady state gap-size $d$, angle $\varphi$ and velocity $u$ as the function of $\beta$ for both the standard squirmer and the  active Janus particle when they move close to the wall. In general, the squirmer and Janus swimmer show similar trends for all the observables $d$, $\varphi$ and $u$ over the $\beta$ range considered (Fig~\ref{beta}). The distance $d$ increases when decreasing $|\beta|$. The largest oscillations (error bars in Fig.~\ref{beta}(A)) are observed near the boundaries between the trapping at the wall and no trapping (dashed lines and dotted lines in Fig.~\ref{beta}(A)).
The biggest difference between the standard squirmer and active Janus particle, is the shift of the wall trapping for smaller absolute values of $\beta$, as can be seen both from the state diagrams (bottom panels in Fig.~\ref{phase_squirmer} and Fig.~\ref{phase_janus} for squirmers and active Janus particles, respectively). The Squirmers show no trapping at the wall for $\beta\in [-4.2, +3.0]$, while for Janus swimmers the unstable region is reduced to $\beta\in[-2.4, +1.9]$.

The observed behaviour of the steady state angles $\varphi$, further strengthens the observation of the similar behaviour between the two types of swimmers (Fig.~\ref{beta}(B)). Both, squirmer and Janus swimmers, show the expected behaviour pushers swimming pointing on average away from the wall while pullers swim in the steady state pointing towards the wall, in agreement with previous simulations and theory~\cite{ignacio2,li14,lintuvuori16,ishimoto13}.
As $\beta$ increases beyond the mode A (Fig.~\ref{phase_squirmer}; squirmer) and I (Fig.~\ref{phase_janus}; Janus) regimes, the $\varphi$ decreases until the swimmer escapes from the wall ($\beta\approx -4.2$ and $\beta\approx -2.4$ for squirmer and Janus swimmers, respectively).
When the particle escapes the surface, the value of $\varphi$ corresponds to the  reflection angle (regions between two dotted lines (squirmer) or two dashed lines (Janus) in Fig.~\ref{beta}(A)). The escape angle increases with $\beta$ for a pushers and decreases with $\beta$ for a pullers. The maximum reflection angle is observed with a neutral ($\beta=0$) as seen from Fig.~\ref{beta}(B). A puller trapped at the surface show a monotonic  decrease of $\varphi$ when $\beta$ is increased.

Finally we turn to the swimming speed $u$ of the particles near the surface (third row in Fig.~\ref{phase_squirmer} and Fig.~\ref{phase_janus} as well as Fig.~\ref{beta}(C)). In all the cases where the swimmers are trapped near the surface, an increase of the swimming velocity $u$ as compared to the free bulk swimming speed $u_0$ is observed. This can be up to $u/u_0\sim 2.5$ for high $|\beta|$ values, as seen from Fig.~\ref{beta}(C). The trends between standard squirmer and Janus swimmers are very similar (circles and squares in Fig.~\ref{beta}, respectively): a linear increase of $u/u_0$ is observed when the absolute value of the squirming parameter $|\beta|$ is increased. When the particle escapes the wall, $u/u_0\approx 1$ is recovered as required.


\begin{figure}
\includegraphics[width=0.5\textwidth]{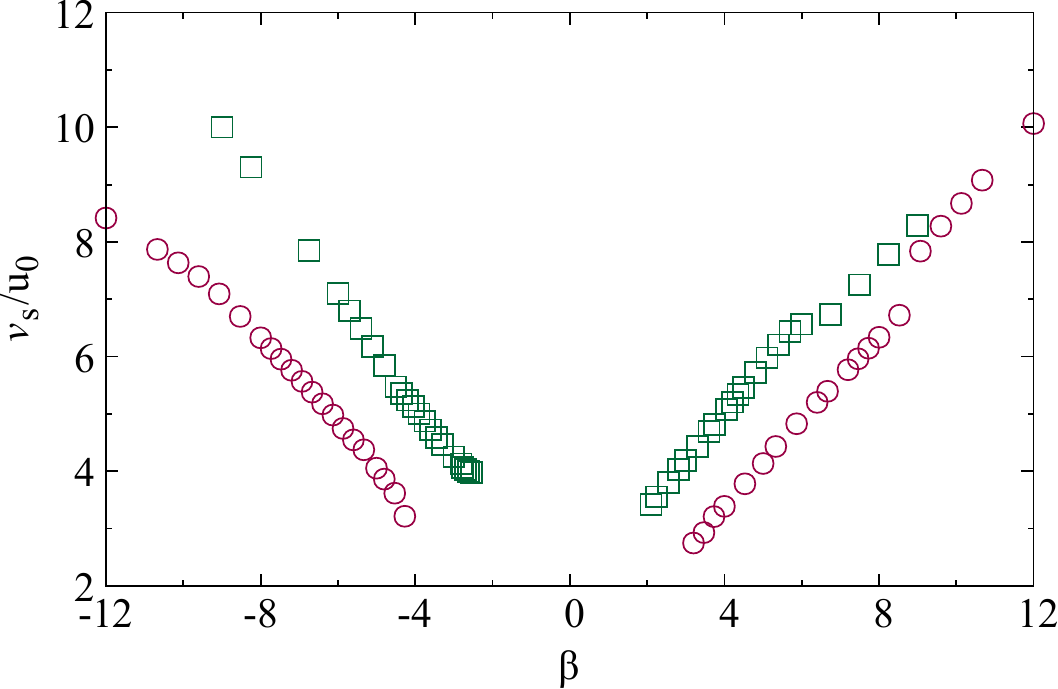}
\hspace*{5cm}       
\caption{The surface velocity $v_s = v_s(\theta_{\text{contact}})$ on the particle surface at the point closest to the wall as a function of the squirming parameter $\beta$. $v_s(\theta_{\text{contact}})$ is calculated using eqs.~(\ref{janus})~and~(\ref{squirmer}) with $\theta_{\text{contact}}(\beta) = \varphi(\beta) + \pi/2$, where $\varphi(\beta)$ is from Fig.~\ref{beta}(C).}
\label{v_b}       
\end{figure}

The swimming speed of a particle in the bulk is given by the surface average of the slip velocity \cite{anderson89}. A more complex situation occurs for a particle close to a solid surface, since the area near the point of minimum distance plays an important role. Then the speed  can be estimated by considering separately the hydrodynamic drag on the lubrication area and on the remainder of the particle surface. The latter contribution is well approximated by
\begin{equation}
  F_b = 6\pi\eta R(u_{||} - u)
\end{equation}
where $u_\parallel$ is the particle velocity corresponding to the far-field, $u$ the unknown actual speed, and $F_b$ the Stokes drag resulting from their difference.  The lubrication area contributes an additional force
\begin{equation}
  F_I = \lambda 6\pi\eta R(v_{s} - u),
\end{equation}
which is proportional to the difference between swimming speed and the slip velocity at minimum distance, $v_s(\theta_{\rm{contact}})$. The prefactor $\lambda = \lambda_0 +\lambda_1\ln(R/d)$ consists of a small constant and a logarithmic correction which arises from singular perturbation schemes in lubrication problems \cite {happel83}. Since there is no net external force, $F_b + F_I = 0$, one obtains the particle velocity
\begin{equation}
  u=\frac{u_{||} + \lambda v_s}{1+\lambda}.\label{speed}
\end{equation}
The orientational angle $\varphi$ can in principle determined from a similar relation for the torques $T_b + T_I = 0$.


The contribution from the far-field hydrodynamics $u_{||}$ can be estimated in the presence of a no-slip surface as $u_{||}\sim -\beta\sin(2\varphi)$~\cite{spagnolie12,schaar15,lintuvuori16}. In the steady state $\varphi < 0$ ($\varphi > 0$) for $\beta > 0$ puller ($\beta < 0$ pusher) (Fig.~\ref{beta}(C)), thus one arrives to a linear relation $u_{||}\sim |\beta|$.  The surface slip velocity at contact $u_s$ includes the contribution both from the steady state swimming angle $\varphi$ (Fig.~\ref{beta}(B)) and the surface flows (equations (\ref{janus}) and (\ref{squirmer}) for the surface slip velocity of the Janus particle and squirmer, respectively). Using the angle data from Fig.~\ref{beta}(B) and identifying that the angle between the swimmers direction and point on the particle surface closest to the wall is given by $\theta_{\text{contact}}(\beta) = \varphi(\beta) +\pi/2$ one can calculate the $v_s(\beta)=u(\theta_{\text{contact}})$ using eq.~(\ref{janus}) and (\ref{squirmer}).  In Fig.~\ref{v_b} $v_s$ is shown for the $\beta$ values where a trapping of the swimmers at the surface was observed. Comparing the $v_s$ and the observed swimming speed $u$ (Fig.~\ref{v_b} and Fig.~\ref{beta}(C)) one can see that $v_s>>u$, implying that $\lambda << 1$ in the equation (\ref{speed}). Interestingly both $v_s$ and $u$ show a linear scaling $v_s,u\sim |\beta|$, for both the Janus swimmers and squirmers (Fig.~\ref{v_b} and Fig.~\ref{beta}(C)).

\section{Conclusions}
In this work we have studied the hydrodynamics of an active spherical particle near a no-slip surface. We have developed a model for an active Janus particle. This corresponds to experimental situation where artificial micro-swimmers can be realised by chemically or thermally driven Janus particles. Typically they consists of colloidal spheres with active cap covering half of their surface. The particle motion results from a finite slip velocity, which takes different values on the two hemispheres due to the different material properties. This renders the slip velocity discontinuous at the particle's equator. We provide a straight forward mapping between the active Janus particle to standard squirmer model, and study the hydrodynamics of these two models via lattice Boltzmann (LB) simulations. Our results show that the far-field hydrodynamics between these two models are very similar, despite that the surface slip velocities (and thus near-field) differ. In order to study the near-field effects, we carried simulations probing the hydrodynamic interactions of the active particle near a no-slip surface. Generally the two models behave similarly over large range of the squirming parameter $\beta$. In both cases we could identify 6 distinct swimming modes of the particles trapped by the boundary. Some key differences arise as well. Notably, at low $|\beta|$ we observe that the affinity of the particle to be trapped by the surface is increased for the Janus swimmers, when compared to the standard squirmer model. Finally we find that when the particles are trapped by the wall, their swimming speed is increased as compared to bulk, as has been seen for example in experiments of colloidal swimmers trapped at a fluid interface~\cite{wang15}. Interestingly our simulation data implies a linear relation between the swimming speed and the magnitude of the squirming parameter $u\sim |\beta|$, for both active Janus particles and squirmers trapped at the boundary.

%
%
\section{Authors contributions}
All the authors were involved in the preparation of the manuscript.
All the authors have read and approved the final manuscript.

\section{Acknowledgements}
Z.S. and J.S.L. acknowledges support by IdEx (Initiative d'Excellence) Bordeaux and computational resources from Avakas cluster. A.W. acknowledges support by the French National Research Agency through
Contract No. ANR-13-IS04-0003.

%
\bibliographystyle{unsrt}
\bibliography{REFERENCELIST}
%
%
%


\end{document}